\documentclass{article}

\usepackage{mathtools}
\usepackage{amssymb}
\usepackage{bbold}
\usepackage[toc, page]{appendix}
\usepackage{color}
\usepackage{graphicx}
\usepackage[a4paper, left=4cm]{geometry}
\usepackage{hyperref}
\usepackage{cleveref}
\usepackage{setspace}
\usepackage{epstopdf}
\usepackage{authblk}
\usepackage{titlesec}

\titleformat{\section}{\normalfont\bfseries}{\thesection.}{0.5em}{}
\titleformat{\subsection}{\normalfont\bfseries}{\thesubsection.}{0.5em}{}
\titleformat{\subsubsection}{\normalfont\bfseries}{\thesubsubsection}{0.5em}{}

\numberwithin{equation}{section}
\numberwithin{figure}{section}

\crefname{appsec}{appendix}{appendices}

\newcommand{\ii}{\text{i}}
\newcommand{\ee}{\text{e}}

\newcommand{\curPath}{.}

\newcommand{\pd}[2]{\frac{\partial #1}{\partial #2}}
\newcommand{\pdn}[3]{\frac{\partial^{#3} #1}{\left(\partial #2\right)^{#3}}}

\newcounter{cmCounter}
\allowdisplaybreaks

\title{Aspects of integrability in a classical model for non-interacting fermionic fields}
\author[1]{S. Grosse-Holz}
\author[1]{T. Engl}
\author[1]{K. Richter}
\author[1]{J. Urbina\thanks{Juan-Diego.Urbina@physik.uni-regensburg.de}}
\affil[1]{Institut f\"ur Theoretische Physik, Universit\"at Regensburg, D-93040 Regensburg, Germany}

\begin{document}

\maketitle

\abstract{In this work we investigate the issue of integrability in a classical model for non-interacting fermionic fields. This model is constructed via classical-quantum correspondence obtained from the semi-classical treatment of the quantum system. Our main finding is that the classical system, contrary to the quantum system, is not integrable in general. Regarding this contrast it is clear that in general classical models for fermionic quantum systems have to be handled with care. Further numerical investigation of the system showed that there may be islands of stability in the phase space. We also investigated a similar model that is used in theoretical chemistry and found this one to be most probably integrable, although also here the integrability is not assured by the quantum-classical correspondence principle.}

\vspace{2ex}

\clearpage
\tableofcontents
\clearpage

\section{Introduction}

The semi-classical approximation in quantum mechanics is by now an established tool to study in a comprehensible and pictorial, but still quantitative manner, non-classical characteristics of the quantum domain, in particular coherent phenomena due to quantum interference \cite{Gutzwiller,Richter}.

The semi-classical program is in nature quite different from quasi-classical approaches where the classical limit of the quantum theory is explicitly used in the form of trajectories that mark the evolution of expectation values of, for example, position and momentum. The regime of validity of the quasi-classical approximation is given by the possibility of reconstructing, at least in an approximate way, the quantum state using only its average position and momentum. A typical example is the time-dependent evolution of well localized, gaussian wavepackets \cite{Sakurai}. Starting with the initial time, and assuming that the wavepacket essentially does not disperse, one can simply use the solutions of the \emph{initial value} classical problem to guide the wavepacket's evolution. It is apparent that the approximation fails when either the wavepacket cannot be adequately represented by a well localized Gaussian, or when different pieces of the initial distribution evolve in a non-rigid way and get to interfere with each other. In closed systems the time scale where these effects arise is called the Ehrenfest time, making reference to the short-time regime where the Ehrenfest theorem can be used to approximate quantum evolution by classical dynamics.

The starting point of the semi-classical approach is, on the contrary, purely quantum-mechanical, namely, the path integral formulation \cite{Schulman}. One extremely important feature of the path integral picture of quantum mechanics is that it allows us to \emph{define} the classical limit of the theory. Contrary to the quasi-classical approach, the semi-classical methods do not require that the classical limit is in any way "correct". From the point of view of the Feynman formulation, a classical limit is defined by the functional that appears as exponent in the integral over paths.

The semi-classical approximation simply amounts for the evaluation of the path integral by the method of stationary phase (SPA) \cite{Gutzwiller,Brack}. It is of course satisfactory that the paths selected by the SPA are precisely the ones that satisfy Hamilton's principle of least action, and therefore are "classical". However, one must keep in mind that classical mechanics, from this point of view, is nothing but a set of rules that tell us how to calculate actions that appear as exponents in semi-classical propagators.

This tension between what one's intuition indicates as classical limit and what the path integral demands to be used in order to construct semi-classical propagators reaches a maximum if the quantum theory we attempt to describe is a fermionic quantum field \cite{Negele}. Here we must try to think classically about extremely counter-intuitive quantum manifestations of the kinematics of fermions, in particular the Pauli exclusion principle and the antisymmetry of the fermionic states. Nevertheless, a semi-classical approach to systems of interacting fermions is of utmost importance in several branches of physics, ranging from quantum chemistry to cosmology.

Several approaches have been proposed to provide a classical theory that gives a glimpse of the dynamics of fermionic fields \cite{Voorhis}, but to our knowledge the only one that actually starts with the fermionic path integral and ends up with the associated van Vleck-Gutzwiller semi-classical propagator by strict application of the SPA was presented only recently \cite{PhDTom}. The huge advantage of this approach is that one can be sure that, whatever strange or counter-intuitive features the classical limit has, it provides the unique consistent way to use a classical theory to study quantum interference effects for fermionic quantum fields through the systematic use of the semi-classical approximation.

In this thesis, one of these apparently contradictory properties of the classical limit for fermionic fields is studied, namely the curious fact that, due to the intrinsic non-linearity of the saturation phenomena associated with the Pauli-principle, the classical limit of the free (non-interacting) fermionic field turns out to be non-integrable in general. This is in sharp  contrast with the bosonic case \cite{PhDTom}, where non-interacting fields are paradigmatic examples of quantum systems with classical integrable limit.

\section{Outline of the quantum mechanical system}
Consider a system of $L$ (bosonic or fermionic) spinless single particle states $|i\rangle,\, i\in \lbrace 1,\ldots,L\rbrace$ (for example lattice sites) with eigenenergies $h_{ii}$ and hoppings $h_{ij}$ (where $h_{ij} = h_{ji}^*$ to ensure hermiticity). Then the Hamiltonian of the system (in second quantization) is given by
\begin{equation}
\hat{H}=\sum_{i,j=1}^L h_{ij} \hat{a}_i^\dagger \hat{a}_j
\end{equation}
where $\hat{a}_i^\dagger$ and $\hat{a}_i$ are the creation and annihilation operators for the state $i$ satisfying
\begin{equation}
\left[\hat{a}_i, \hat{a}_j^\dagger\right]_\pm = \delta_{ij}\, ,\quad \left[\hat{a}_i, \hat{a}_j\right]_\pm=0\, ,
\end{equation}
where $[A, B]_- := AB - BA$ is the commutator, which is used if the $\hat{a}_i$ are of bosonic character and $[A,B]_+ := AB + BA$ is the anticommutator, used for fermions.

As the matrix $\mathbf{h}$ is hermitian, we can find a unitary similarity matrix $\mathbf{u}$ to diagonalize it. We then can use the matrix $\mathbf{u}$ to transform the creation and annihilation operators:
\begin{equation}
\begin{matrix}\hat{a}_i = \sum_{k=1}^L u_{ik}\hat{A}_k \\\\ \hat{a}_i^\dagger = \sum_{k=1}^L u_{ik}^*\hat{A}_k^\dagger \end{matrix} \Longleftrightarrow \begin{matrix}\hat{A}_k = \sum_{i=1}^L u_{ik}^*\hat{a}_i \\\\ \hat{A}_k^\dagger = \sum_{i=1}^L u_{ik}\hat{a}_i^\dagger \end{matrix}
\end{equation}
where we used the unitarity of $\mathbf{u}$: $(\mathbf{u}^{-1})_{ij}=u_{ji}^*$.

As the transformation is linear, the (anti-)commutation relations of the creation and annihilation operators are preserved:
\begin{align}
\left[\hat{A}_k, \hat{A}_l^\dagger \right]_\pm =&\, \sum_{k,l=1}^L u_{ik}^* u_{jl} \left[\hat{a}_k, \hat{a}_l^\dagger \right]_\pm = \sum_{k,l=1}^L u_{ik}^* u_{jl} \delta_{kl} = \left(\mathbf{u}^{-1}\mathbf{u}\right)_{ij} = \delta_{ij}\,, \label{eq:commAA1}\\
\left[\hat{A}_k, \hat{A}_l \right]_\pm =&\, \sum_{k,l=1}^L u_{ik}^* u_{jl} \left[\hat{a}_k, \hat{a}_l \right]_\pm = 0\,. \label{eq:commAA2}
\end{align}
Using this transformation, $\hat{H}$ becomes diagonal:
\begin{equation}
\hat{H} = \sum_{i,j,k,l=1}^L h_{ij}u_{ik}^*u_{jl}\hat{A}_k^\dagger \hat{A}_l = \sum_{k=1}^L w_k \hat{A}_k^\dagger \hat{A}_k \equiv \sum_{k=1}^L w_k\hat{N}_k
\end{equation}
where $w_k = \sum_{i,j=1}^L u_{ik}^* h_{ij} u_{jk}$ are the eigenvalues of $\mathbf{h}$.

Because of \cref{eq:commAA1,eq:commAA2}, the $\hat{N}_k$ commute:
\begin{align}
\left[\hat{N}_k, \hat{N}_l\right] =&\, \hat{A}_k^\dagger \hat{A}_k \hat{A}_l^\dagger \hat{A}_l - \hat{A}_l^\dagger \hat{A}_l \hat{A}_k^\dagger \hat{A}_k \notag\\
=&\, \hat{A}_k^\dagger \hat{A}_l \delta_{kl} \pm \hat{A}_k^\dagger \hat{A}_l^\dagger \hat{A}_k \hat{A}_l - (\hat{A}_l^\dagger \hat{A}_k \delta_{kl} \pm \hat{A}_l^\dagger \hat{A}_k^\dagger \hat{A}_l \hat{A}_k) = 0
\end{align}
where ``$+$'' is used in case of bosons and ``$-$'' in case of fermions. Because of this, the $\hat{N}_k$ obviously also commute with $\hat{H}=\sum_{k=1}^L w_k \hat{N}_k$, thus we have found $L$ independent constants of motion. This means the system is quantum-integrable.

In the original basis, the constants of motion read
\begin{equation} \label{eq:comQM}
\hat{N}_k = \hat{A}_k^\dagger \hat{A}_k = \sum_{i,j=1}^L u_{jk}^* u_{ik} \hat{a}_i^\dagger \hat{a}_j\,.
\end{equation}
If the quantum-classical correspondence works well, we can expect the clasical system to behave in the same manners as the quantum one. Especially, it should also be integrable.

\section{Bosons} \label{sec:Bosons}
The classical analogy in the bosonic case is very simple: the corresponding classical Hamiltonian for the bosonic case is obtained by replacing $\hat{b}_i$ and $\hat{b}_i^\dagger$ by complex fields $\psi_i$ and $\psi_i^*$ respectively \cite{PhDTom,engl2014coherent}. This may only be done for normally ordered products of creation and annihilation operators (which means all creation operators are to the left of all annihilation operators), so there is no ordering ambiguity.

Given this, we can write down the classical Hamiltonian in the bosonic case:
\begin{equation} \label{eq:BosH}
H = \sum_{i,j=1}^L h_{ij} \psi_i^* \psi_j
\end{equation}
By the classical-quantum correspondence we expect conservation (of classical versions) of the quantities \labelcref{eq:comQM}:
\begin{equation} \label{eq:COMBos}
N_k = \sum_{i,j=1}^L u_{jk}^* u_{ik} \psi_i^* \psi_j
\end{equation}
It has to be emphasized that these functions are just an educated guess for the constants of motion of the classical system \labelcref{eq:BosH}.

However, it can be shown by direct calculation, that this guess indeed gives $L$ independent constants of motion, meaning that the classical system, in perfect analogy to the quantum system, is integrable.

\section{Fermions}
As in the bosonic case (\cref{sec:Bosons}), the classical system is obtained by replacing the operators with complex functions. But because of the anticommutativity (versus the commutativity of the complex fields $\psi$ and $\psi^*$), the replacements are more complex. In~\cite{PhDTom} they are given as
\begin{align}
\hat{c}_i^\dagger \hat{c}_i \rightarrow &\, |\psi_i|^2\,,\\
\hat{c}_i^\dagger \hat{c}_{j\neq i} \rightarrow &\, \psi_i^* \psi_j \ee^{-|\psi_i|^2-|\psi_j|^2} \prod_{k=\min(i, j) + 1}^{\max(i, j)-1}(1-2|\psi_k|^2)\,.
\end{align}
Here, however let us consider the slightly more general version
\begin{align}
\hat{c}_i^\dagger \hat{c}_i \rightarrow &\, |\psi_i|^2\,, \label{eq:L2replace1}\\
\hat{c}_i^\dagger \hat{c}_{j\neq i} \rightarrow &\, \psi_i^* \psi_j f(|\psi_i|^2)f(|\psi_j|^2) \prod_{k=\min(i, j) + 1}^{\max(i, j)-1}(1-2|\psi_k|^2) \label{eq:L2replace2}
\end{align}
with an arbitrary smooth function $f(x)$.

Of course $f(x)$ in reality is not completely arbitrary. For example setting $f(x) \equiv 0$ obviously would not be a reasonable choice. The limits of the freedom in the choice of $f$ are investigated in \cite{engl2014semiclassical}. The two most important choices are $f(x) = \ee^{-x}$ and $f(x) = \sqrt{1-x}$.

Due to these more complicated replacements, it is not possible (with affordable effort) to compute the Poisson-brackets for a Hamiltonian with a general number $L$ of sites. So we start with the first non-trivial case, which is $L=2$.

\subsection{Two sites} \label{subsec:L2}
With a fixed number of sites given, we can explicitly write down the matrix $\mathbf{h}$ determining the Hamiltonian:
\begin{equation}
\mathbf{h} = \begin{pmatrix} \epsilon_1 & J \\ J^* & \epsilon_2 \end{pmatrix}
\end{equation}
This form (with $\epsilon_1, \epsilon_2 \in \mathbb{R}$) is required by the hermiticity of $\mathbf{h}$.

Using the replacement rules \labelcref{eq:L2replace1,eq:L2replace2} we get
\begin{align} \label{eq:H2rep}
H =&\, \epsilon_1 |\psi_1|^2 + \epsilon_2 |\psi_2|^2 + (J\psi_1^*\psi_2 + J^*\psi_1\psi_2^*)f(|\psi_1|^2)f(|\psi_2|^2) \,,\\
N_k =&\, |u_{1k}|^2|\psi_1|^2 + |u_{2k}|^2|\psi_2|^2 + (u_{1k}u_{2k}^* \psi_1^*\psi_2 + u_{1k}^*u_{2k}\psi_1\psi_2^*)f(|\psi_1|^2)f(|\psi_2|^2) \,.\label{eq:N2rep}
\end{align}
Given this it can be shown by explicit calculation, that the relevant Poisson-brackets vanish:
\begin{equation}
\lbrace H, N_k \rbrace = 0 \,\forall k\in\lbrace 1, 2\rbrace,\, f\in C^1(\mathbb{R})
\end{equation}
and
\begin{equation}
\lbrace N_k, N_l \rbrace = 0 \,\forall k, l\in\lbrace 1, 2\rbrace,\, f\in C^1(\mathbb{R}).
\end{equation}
This implies that for $L=2$, the classical analogy to the fermionic quantum system is integrable. So now let us continue with $L=3$.

\subsection{Three sites} \label{subsec:L3}

For the case $L=3$ calculations similar to those in \cref{subsec:L2} can be done. We will consider two different (yet very similar) systems: a linear one, where we only consider the interactions $|1\rangle \leftrightarrow |2\rangle$ and $|2\rangle \leftrightarrow |3\rangle$, and a circular system, where also the states $|1\rangle$ and $|3\rangle$ can interact. For simplicity we assume the same strength for all interactions.

The linear Hamiltonian is given by the matrix
\begin{equation} \label{eq:hstring}
\mathbf{h}=\begin{pmatrix} \epsilon_1 & J & 0 \\ J^* & \epsilon_2 & J \\ 0 & J^* & \epsilon_3 \end{pmatrix}
\end{equation}
and for the cyclic sytem we get
\begin{equation} \label{eq:hcirc}
\mathbf{h}=\begin{pmatrix} \epsilon_1 & J & J \\ J^* & \epsilon_2 & J \\ J^* & J^* & \epsilon_3 \end{pmatrix}\,.
\end{equation}
The calculations for the two cases are very similar and especially yield equivalent results, so here it shall suffice to discuss just the (shorter) one for \cref{eq:hstring} in detail.

\subsubsection{Results}\label{subsubsec:resL3}

The full calculations are very cumbersome, thus here we shall just state the result:
\begin{align} \label{eq:resL3}
\lbrace H, N_k \rbrace = 2\ii\Im\Big{(}
&\,- 2Ju_{1k}u_{3k}^*\psi_1^*\psi_1^*\psi_3\psi_2 f^2(|\psi_1|^2)f(|\psi_3|^2)f(|\psi_2|^2) \notag\\
&\,- 2J^*u_{1k}^*u_{3k}\psi_1\psi_3^*\psi_3^*\psi_2 f(|\psi_1|^2)f^2(|\psi_3|^2)f(|\psi_2|^2) \notag\\
&\, + Ju_{1k}^*u_{3k}\psi_1\psi_2^*\big( (1-2|\psi_2|^2) g(|\psi_3|^2) - 1 + 2|\psi_3|^2f^2(|\psi_3|^2)\big)f(|\psi_1|^2)f(|\psi_2|^2) \notag\\
&\, + J^*u_{1k}u_{3k}^*\psi_3\psi_2^*\big((1-2|\psi_2|^2) g(|\psi_1|^2) - 1 + 2|\psi_1|^2f^2(|\psi_1|^2) \big)f(|\psi_3|^2)f(|\psi_2|^2) \notag\\
&\, + (Ju_{3k}^*u_{2k}\psi_1^*\psi_3 + J^*u_{1k}^*u_{2k}\psi_1\psi_3^*)\big(g(|\psi_2|^2) - (1-2|\psi_2|^2)\big)f(|\psi_1|^2)f(|\psi_3|^2) \Big{)}\,.
\end{align}
Now the question is: can we find a function $f(x)$, such that this vanishes identically? The important part of the answer is: no. However, there are special cases (e.g. $J=0$) where the Poisson-brackets do vanish, so we should investigate this issue in more detail.

This detailed investigation is rather cumbersome and technical, thus here we shall just quote the result. The full considerations can be found in \cref{app:tech}.

Ultimately, we find the following: the right hand side of \cref{eq:resL3} can only vanish identically if
\begin{equation} \label{eq:vanish}
Ju_{1k}u_{3k}^*\psi_1^*\psi_1^*\psi_3\psi_2 f^2(\Psi_1^2)f(\Psi_3^2)f(\Psi_2^2) = 0\,.
\end{equation}
However, reasonable assumptions (\cref{subsubsec:tech}) state that every factor in this term is non-zero.

\subsubsection{Discussion}

In \cref{subsubsec:resL3} we stated that under reasonable assumptions (\cref{subsubsec:tech}) the Poisson-brackets \labelcref{eq:resL3} cannot vanish identically. Thus there are two possibilities: either the Poisson-brackets really do not vanish or the assumptions were wrong. In this paragraph we will investigate this issue and see that both possibilities are partly true. In fact, the assumptions are legit for any generic setup, but there are special cases where we may not use them. In most of these special cases, it is very easy to see that the Poisson-brackets do vanish.

The most important point is that the classical-quantum correspondence does not give full freedom of choice for $f$. Rather there are very restrictive conditions to be met, which essentially exclude pathological examples. Thus in \cref{subsubsec:tech,subsubsec:compPois} we could innocently assume that everything is ``nice'', e.g. there are points in phase space where the $f$'s and $|\psi|$'s are non-zero.

To be precise, the most important example is $f(x) = \ee^{-x}$ which certainly fulfills the non-zero conditions by vanishing nowhere in the phase space. Thus the proof certainly holds for this case.

Especially in the field of theoretical chemistry, $f(x) = \sqrt{1-x}$ is widely used. This provides certain particularities. First of all, the domain of $f$ is bounded to $x \leq 1$. Thus we now have a boundary of the domain, which we have to consider separately. Assume we start with the system in a configuration where $|\psi_i|^2 = 1\,\forall i$. As shown below (\cref{eq:comN,eq:NPsi}), the quantity $\sum_i |\psi_i|^2$ is a conserved quantity. But because of the condition $|\psi_i|^2 \leq 1$, this means that the $|\psi_i|^2$'s can neither increase nor decrease, because as one of them got smaller, another one would have to grow, which is not possible. Thus $|\psi_i|^2 = \text{const.} = 1\,\forall i$, which directly implies $f(|\psi_i|^2) = \text{const.} = 0\,\forall i$. Thus the Poisson-brackets do vanish on any trajectory that runs on the border of the domain.

However, besides this special case all the conditions also apply for $f(x) = \sqrt{1-x}$, so inside the domain the Poisson-brackets still do not vanish.

\subsubsection{Consequences}

Now we shall discuss the implications of the fact that the Poisson-brackets \labelcref{eq:resL3} do not vanish. The bottom line is that for $L=3$, the $N_k$ are in general no longer constants of motion. This does not directly imply that the system is not integrable anymore, but our educated guess for the constants of motion is just wrong.

But still the $N_k$ are not completely useless. The interpretation of $\hat{N}_k$ in the quantum system is the number of particles in state $k$. Of course, the overall number of particles $\hat{N}=\sum_{k=1}^L \hat{N}_k$ in the system is conserved, and this stays true in the classical analogon. To show this, we note that all the terms in \cref{eq:resL3} are proportional to a combination of $u_{ij}$'s of the form $u_{ik}u_{jk}^*$ where $i\neq j$. Employing the unitarity of $\mathbf{u}$, these expressions, when summed over $k$, yield $\delta_{ij}$ which is zero for $i\neq j$. Thus
\begin{equation}\label{eq:comN}
\lbrace H, N \rbrace = \sum_{k=1}^3 \lbrace H, N_k \rbrace = 0\,,
\end{equation}
so the total particle number, as stated above, still is conserved. It can be shown that
\begin{equation}\label{eq:NPsi}
N = |\psi_1|^2 + |\psi_2|^2 + |\psi_3|^2\,.
\end{equation}

The last thing to mention about \cref{eq:resL3} is that all terms are proportional to $J$ or $J^*$, so a system of three independent fermionic states (which means $J=0$) still is integrable, which is also trivially seen from its separability.

\section{Numerical investigation of the fermionic L=3 system}\label{sec:numerics}
From the results of \cref{subsec:L3} we can make no statement concerning the integrability of the system in general. To investigate this issue, we simulated the system and looked at Poincar\'e-surfaces of section. In order to do so, we first transform the canonical variables in a suitable way which is discussed in \cref{subsec:physTrafos}. The numerical results are shown in \cref{subsec:images}.

\subsection{Variable transformations}\label{subsec:physTrafos}
In the above considerations we saw that for $L=3$ the direct analogy between the quantum and the classical system does not give enough constants of motion to ensure integrability anymore. However, there still are constants of motion: of course $H$ itself, as it is time-independent, is a constant of motion and in \cref{eq:comN,eq:NPsi} we saw that the total particle number
\begin{equation}
N = N_1 +N_2 +N_3 = |\psi_1|^2 + |\psi_2|^2 + |\psi_3|^2
\end{equation}
also is a constant of motion. Using this, we can reduce the dimensionality of our problem by variable transformations.

First of all, we want to get rid of the complex variables $\psi_i$ and $\psi_i^*$ and instead express our system by means of real variables. Doing so we have to take care about the fact that the equations of motion for $\psi_i$ and $\psi_i^*$ (which are given for example in~\cite{PhDTom} \namecrefs{eq:resL3} (4.2a,b)) are not the standard Hamilton's equations. Instead there is an additional factor $\ii$ in them (which is needed to assure consistency). But for the real variables we are about to introduce we must have Hamilton's equations without this prefactor. Thus the first transformation introduced here cannot be canonical, as it will not precisely preserve the equations of motion.

However, we can get close to the needed transformation by considering the generating function $F=\sum_{i=1}^3 \frac{1}{2}\psi_i^2\ee^{-2\ii \theta_i}$ which yields the (canonical) transformation
\begin{equation}
\psi_i^* = \pd{F}{\psi_i} = \psi \ee^{-2\ii\theta}\,,\quad \tilde{n}_i = \pd{F}{\theta_i} = -\ii \psi_i^2 \ee^{-2\ii\theta} = -\ii \psi_i\psi_i^*\,.
\end{equation}
Now we introduce the coordinates $n_i := \ii \tilde{n}_i = |\psi_i|^2$. This cancels the $\ii$ in the equations of motion, so in the real variables we obtain Hamilton's equations
\begin{equation}
\dot{n_i} = \pd{H}{\theta}\,,\quad \dot{\theta_i} = -\pd{H}{n_i}\,.
\end{equation}
From now on we will consider the $\mathbf{h}$ given in \cref{eq:hcirc}. Using the new coordinates $n_i$ and $\theta_i$, the Hamiltonian then reads
\begin{align} \label{eq:Hntheta}
H =&\, n_1\epsilon_1 + n_2\epsilon_2 + n_3\epsilon_3 \notag\\
&\,+ (J\ee^{-\ii\theta_1 +\ii\theta_2} + J^*\ee^{\ii\theta_1 - \ii\theta_2})\sqrt{n_1 n_2}\,f(n_1)f(n_2) \notag\\
&\,+ (J\ee^{-\ii\theta_2 + \ii\theta_3} + J^*\ee^{\ii\theta_2 - \ii\theta_3})\sqrt{n_2n_3}\,f(n_3)f(n_2) \notag\\
&\,+ (J\ee^{-\ii\theta_1 + \ii\theta_3} + J^*\ee^{\ii\theta_1 - \ii\theta_3})(1-2n_2)\sqrt{n_1n_3}\,f(n_1)f(n_3)\,.
\end{align}
Note that \cref{eq:NPsi} in these variables reads
\begin{equation}
N = n_1 + n_2 + n_3\,.
\end{equation}
Now we will exploit the conservation of $N$ to reduce the dimensionality of the problem. Therefore we again introduce new coordinates $((n,\alpha),(m,\beta),(N,\Theta))$ where we want $N$ to be the total particle number. We transform to these coordinates by means of the generating function
\begin{equation}
F(n,m,N,\theta_1,\theta_2,\theta_3) = n(\theta_1-\theta_2) + m(\theta_3 - \theta_2) + N\theta_2\,,
\end{equation}
which yields the transformations
\begin{equation}
\begin{matrix}
\alpha = \pd{F}{n} = \theta_1 - \theta_2\,, & \beta = \pd{F}{m} = \theta_3 - \theta_2\,, & \Theta = \pd{F}{N} = \theta_2\,, \\&&&\\
n_1 = \pd{F}{\theta_1} = n\,, & n_2 = \pd{F}{\theta_2} = N - n_1 - n_3\,, & n_3 = \pd{F}{\theta_3} = m\,.
\end{matrix}
\end{equation}
This trivially gives $N=n_1+n_2+n_3$ which is what we intended to do.

Rewriting the Hamiltonian \labelcref{eq:Hntheta} in these coordinates, we obtain
\begin{align} \label{eq:Hnalpha}
H =&\, n(\epsilon_1-\epsilon_2) + N\epsilon_2 + m(\epsilon_3-\epsilon_2) \notag\\
&\,+ (J\ee^{-\ii\alpha} + J^*\ee^{\ii\alpha})\sqrt{n}\sqrt{N-n-m}\,f(n)f(N-n-m) \notag\\
&\,+ (J\ee^{\ii\beta} + J^*\ee^{-\ii\beta})\sqrt{m}\sqrt{N-n-m}\,f(m)f(N-n-m) \notag\\
&\,+ (J\ee^{-\ii\alpha + \ii\beta} + J^*\ee^{\ii\alpha - \ii\beta})(1-2(N-n-m))\sqrt{nm}\,f(n)f(m)\,.
\end{align}
Note that $H$ does not depend on $\Theta$. So again we see that $N$, as the conjugated momentum to the cyclic coordinate $\Theta$, is conserved. In fact this Hamiltonian can be interpreted as a system with two degrees of freedom (instead of the previous three), where $N$ is just a constant parameter.

The variable pairs $(n, \alpha)$ and $(m, \beta)$ still are of the nature of polar coordinates. For visualization (see below) we want to use coordinates of cartesian nature, thus we perform one last transformation. Unfortunately, the transformation to coordinates $(x_1,x_2,y_1,y_2)$ where $x_1^2 + x_2^2 = n$ and $y_1^2+y_2^2 = m$ is not canonical. Nevertheless we will use these coordinates, which is justified by the following argumentation:

Consider the generating function
\begin{equation}
F = \frac{1}{2}\tilde{x}_1^2 \tan \alpha + \frac{1}{2}\tilde{y}_1^2 \tan \beta\,.
\end{equation}
This ultimately gives the transformations
\begin{equation}
\begin{matrix}
\frac{\tilde{x}_1}{\sqrt{2}} = \sqrt{n}\cos \alpha\,, & \frac{\tilde{x}_2}{\sqrt{2}} = \sqrt{n}\sin \alpha \\&\\
\frac{\tilde{y}_1}{\sqrt{2}} = \sqrt{m}\cos \beta\,, & \frac{\tilde{y}_2}{\sqrt{2}} = \sqrt{m}\sin \beta\,.
\end{matrix}
\end{equation}
so obviously $x_i = \frac{\tilde{x}_i}{\sqrt{2}}$, $y_i = \frac{\tilde{y}_i}{\sqrt{2}}$. However, if we use these coordinates, the equations of motion are no longer the canonical ones. Consider for example the canonical equation of motion for $\tilde{x}_1$ (for brevity we write $\lbrace q_i | i=1,2,3,4 \rbrace = \lbrace x_1, x_2, y_1, y_2 \rbrace$):
\begin{equation}
\dot{\tilde{x}}_1 = \pd{H}{\tilde{x}_2} = \sum_{i=1}^4 \pd{H}{q_i}\pd{q_i}{\tilde{x}_2} = \sum_{i=1}^4 \pd{H}{q_i}\frac{\delta_{i2}}{\sqrt{2}} = \frac{1}{\sqrt{2}}\pd{H}{x_2}\,.
\end{equation}
With $\dot{x}_1 = \frac{\dot{\tilde{x}}_1}{\sqrt{2}}$ this gives $\dot{x}_1 = \frac{1}{2}\pd{H}{x_2}$. Analogous we obtain the factor $\frac{1}{2}$ in every equation of motion. However, this is the only place where the non-canonicity of the coordinate transformation comes into play, and on the other hand there is no other place in the whole system, where the time coordinate appears. Thus we can absorb the constant factor into our definition of time and thus re-obtain the canonical equations of motion. In the end, this just means that the evolution of the system is twice as slow as it would be otherwise, but (as the Hamiltonian is time-independent) the overall speed of the evolution of the system is irrelevant for the evolution itself, which is what we are interested in.

So finally we write the Hamiltonian as
\begin{align} \label{eq:Hxy}
H =&\, (x_1^2+x_2^2)(\epsilon_1-\epsilon_2) + (y_1^2+y_2^2)(\epsilon_3-\epsilon_2) + N\epsilon_2 \notag\\
&\,+ 2\Re(J(x_1-\ii x_2))\sqrt{N-x_1^2-x_2^2-y_1^2-y_2^2}\,f(x_1^2+x_2^2)f(N-x_1^2-x_2^2-y_1^2-y_2^2) \notag\\
&\,+ 2\Re(J(y_1+\ii y_2))\sqrt{N-x_1^2-x_2^2-y_1^2-y_2^2}\,f(x_1^2+x_2^2)f(N-x_1^2-x_2^2-y_1^2-y_2^2) \notag\\
&\,+ 2\Re(J(x_1-\ii x_2)(y_1+\ii y_2))(1-2(N-x_1^2-x_2^2-y_1^2-y_2^2))f(x_1^2+x_2^2)f(y_1^2+y_2^2)\,.
\end{align}

\subsection{Poincar\'e-surfaces of section the fermionic three-site system}

In this section we require some general knowledge of theoretical mechanics and the technique of Poincar\'e-surfaces of section in general. An introduction to this topic can be found for example in \cite{Tabor}.

In order to do a Poincar\'e-surface of section, an important step is to solve the equation $H=E$ for one of the canonical variables. Unfortunately, in our specific considerations, the Hamiltonian \labelcref{eq:Hnalpha} does not have a form which admits a nice solution of this equation for any of the variables. Thus we cannot easily find a parametrization of the energy shell. But we can use an alternate procedure: first we slice the phase space by looking at the coordinate hyperplanes of a canonical variable (now we see why we wanted to transform to coordinates of cartesian character: this better suits the intuitive picture of $\mathbb{R}^n$) and then we manually restrict the image to the energy shell. This restriction is performed by considering only trajectories with a given energy.

\begin{figure}
\includegraphics[width=\textwidth]{\curPath 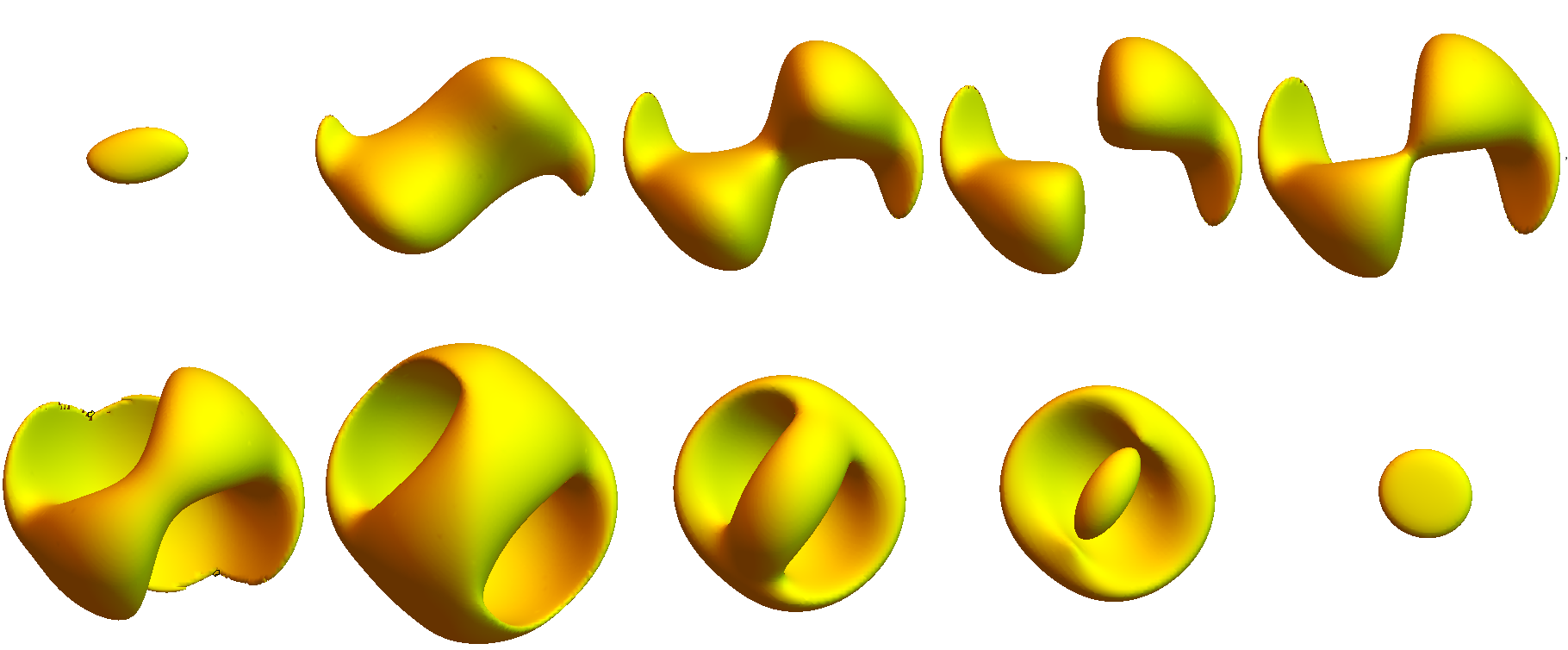}
\caption{Several slices of the energy shell of the Hamiltonian \labelcref{eq:Hnalpha} for a generic choice of parameters. These images are inteded to give an overview over the full (4-dimensional) energy shell. Therefore we fixed the canonical variable $y_1$, thus the three dimensions of every small image represent the remaining three canonical variables. The values for $y_1$ increase from top-left to bottom-right in non-equidistant steps.}
\label{fig:eshell}
\end{figure}

To get familiar with the energy shell of our concrete problem, an example is depicted in \cref{fig:eshell}.

\subsection{Interpreting Poincar\'e-surfaces of section} \label{subsec:poincareInterpretation}

In this work, the main reason for doing Poincar\'e-surfaces of section is to distinguish between integrable and non-integrable systems. This is very easy, given the following considerations: the system we are looking at has two degrees of freedom and the Hamiltonian is a constant of motion. Thus the question of integrability reduces to the question whether there is another constant of motion or not. If there was a second conserved quantity, the invariant manifolds of the flow of the system would be two-dimensional, thus each of them would appear one-dimensional in the Poincar\'e-surface of section. If, however, there was no second constant of motion, each invariant manifold would be three-dimensional and thus its image in the Poincar\'e-surface of section would be two-dimensional.

Of course the numerics do not directly give pictures of the invariant manifolds of the system, but just distinct points of single trajectories, where each trajectory runs on an invariant manifold. But if the trajectory covers a significant part of the invariant manifold (especially if the trajectory is a dense subset of the manifold), we get a good image of what this manifold looks like. Of course the picture gets better, the longer the trajectories are traced.

So finally the prescription for interpreting the shown Poincar\'e-surfaces of section is as follows: if the points of one trajectory form a (one-dimensional) line, there is a second constant of motion and the system is integrable, whereas if the points spread over a certain (two-dimensional) area, the system is not integrable.

\subsection{Numerical results}\label{subsec:images}

For the numerical calculations we considered the cyclic system (see \cref{subsec:L3,eq:hcirc}). The computations where performed using Wolfram's Mathematica 10.

\begin{figure}
\includegraphics[width=\textwidth]{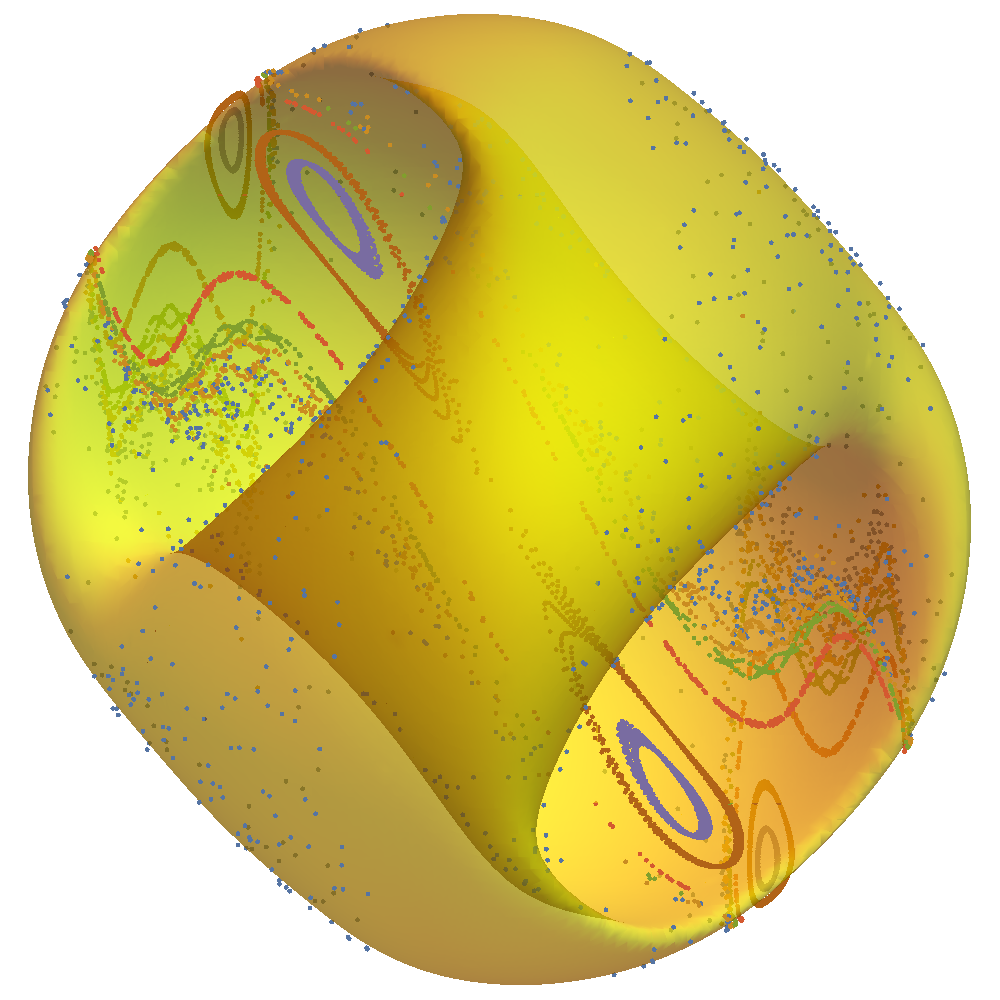}
\caption{Poincar\'e sections of several trajectories (marked by color) of the system \labelcref{eq:Hxy} with $f(x) = \ee^{-x}$ at $\epsilon_1 = \epsilon_2 = \epsilon_3 = 1$, $J=0.6$, $N=3$, $H=3.14$. Apparently there are regions on the energy shell (which is depicted in yellow), where the system displays non-integrable behaviour, whereas it looks (close to) integrable in other regions. The presence of the non-integrable regions proves the result of the calculations in \cref{subsec:L3} where we saw that the educated guess for global constants of motion does not work. Here we see that this is because the system simply is not integrable globally. This implies a partial breakdown of the classical-quantum correspondence. However the ``close to integrable'' regions imply the existence of a local second constant of motion. This is under current investigation.}
\label{fig:Section}
\end{figure}

\begin{figure}
\includegraphics[width=\textwidth]{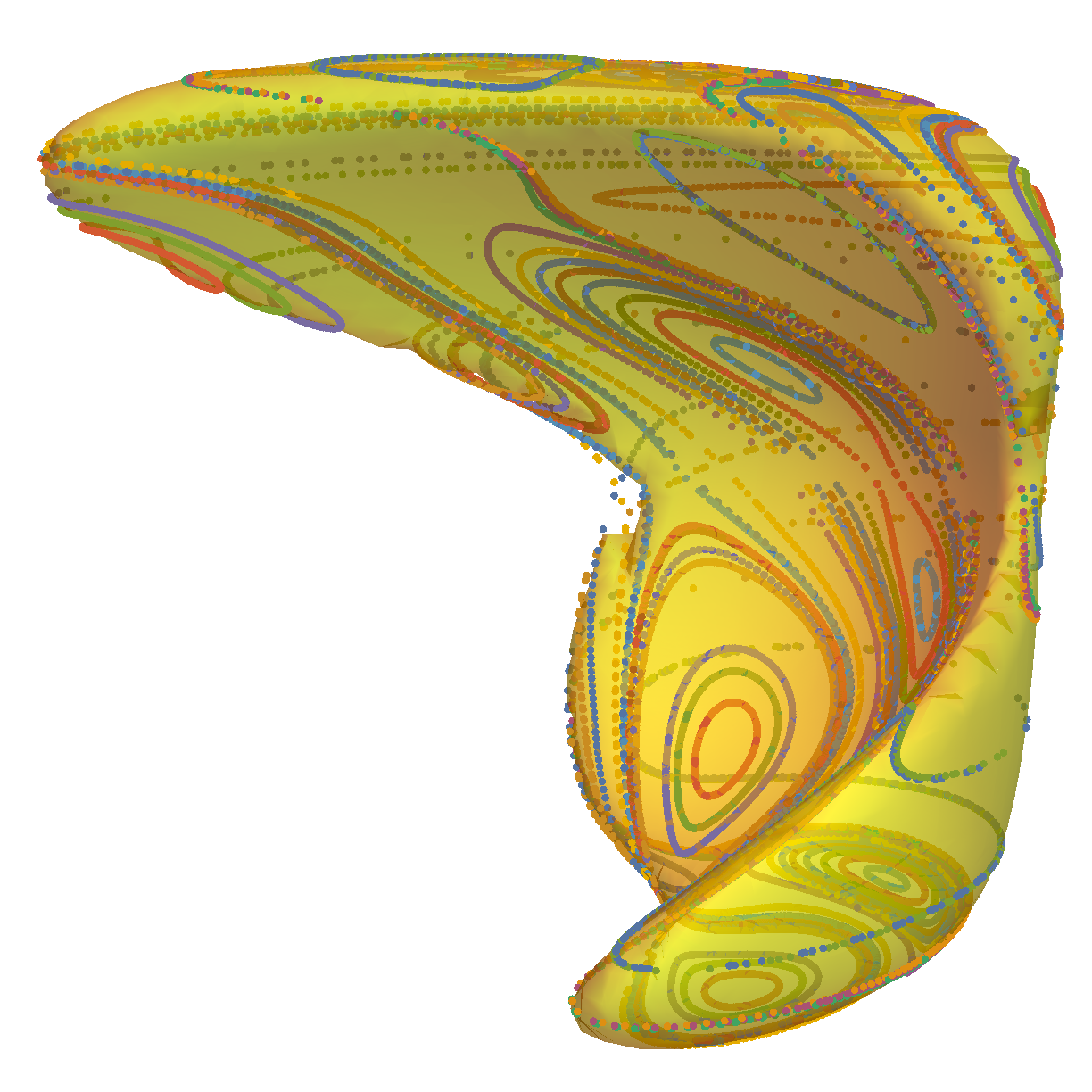}
\caption{Generic Poincar\'e-surface of section of the system \labelcref{eq:Hxy} with $f(x) = \sqrt{1-x}$. According to \cref{subsec:poincareInterpretation} this definitely looks like there should be a second constant of motion in this case. However, direct calculation (\cref{subsec:L3}) shows that the classical-quantum correspondence investigated in this work still does not give enough constants of motion. Thus we conjecture that there is an additional constant of motion which we have no knowledge of.}
\label{fig:chemical}
\end{figure}

The main results of the numerical calculations can be summarized in two pictures, which are \cref{fig:Section} for the case $f(x) = \ee^{-x}$ and \cref{fig:chemical} for the case $f(x) = \sqrt{1-x}$. The implications of these results are explained in the figure captions respectively.

\clearpage

\section{Conclusion}

At the beginning of this work it was shown by direct calculation, that the quantum system under consideration is integrable both in the bosonic and the fermionic case and for arbitrary number of sites.

Using the quantum-classical correspondence from~\cite{PhDTom}, we thus expected the classical Hamiltonian to be integrable. Direct calculations proved that this is the case for the bosonic system, for the fermionic system, however, it was shown that the classical-quantum correspondence in general does not give enough constants of motion.

To further investigate the system, it was simulated numerically, which resulted in the insight that the fermionic three-site system is not integrable in general. This breaks the classical-quantum correspondence. The reasons for this behaviour are subject to current research.

However, the numerics also showed integrable regions in the Poincar\'e-surfaces of section, which indicates an additional (local) constant of motion which we have no knowledge of. In the case of $f(x) = \sqrt{1-x}$ there even seems to be another global constant of motion which would imply that this system is integrable. However, also for $f(x) = \sqrt{1-x}$ we could show that the quantities we guessed from the classical-quantum correspondence are no longer conserved.

\clearpage
\begin{appendices}
\crefalias{section}{appsec}
\section{Investigation of expression \labelcref{eq:resL3}} \label{app:tech}

\subsection{Technicalities}\label{subsubsec:tech}

As they only depend on derivatives of the analytic functions $H$ and $N_k$, the Poisson-brackets are analytic functions in the canonical variables $\psi_i$ and $\psi_i^*$. For an analytic function, to vanish identically is equivalent to the function and all its (partial) derivatives vanishing at some point. So in order to prove that the Poisson-brackets do not vanish identically, it is sufficient to find a point where an arbitrary partial derivative does not vanish.

To exclude trivial solutions, assume that $Ju_{ik}u_{jk}^* \neq 0$ for arbitrary $i \neq j$. Now take a point in phase space where $f(|\psi_i|) \neq 0 \,\forall i$ and $|\psi_i|\neq 0\,\forall i$. It is reasonable to assume $f(x) \neq 0$ on the inside of the domain of $f$, thus such points do exist on any trajectory that does not solely run on the border of the domain of $f$.

Now we have to find suitable partial derivatives. Therefore note that if we define $\psi = \Psi\ee^{\ii\Phi}$ ($\Psi, \Phi\in\mathbb{R}$), we get
\begin{equation}
\pd{f(\psi, \psi^*)}{\Psi} = \pd{f(\psi, \psi^*)}{\psi}\ee^{\ii\Phi} + \pd{f(\psi, \psi^*)}{\psi^*}\ee^{-\ii\Phi}\,,
\end{equation}
\begin{equation}
\pd{f(\psi, \psi^*)}{\Phi} = \pd{f(\psi, \psi^*)}{\psi} \ii \Psi\ee^{\ii\Phi} - \pd{f(\psi, \psi^*)}{\psi^*} \ii \Psi\ee^{-\ii\Phi}\,.
\end{equation}
Thus we can write the Wirtinger derivatives as
\begin{equation}
\pd{}{\psi} = \frac{\ee^{-\ii\Phi}}{2}\left(\pd{}{\Psi} - \frac{\ii}{\Psi}\pd{}{\Phi}\right)\,,\qquad \pd{}{\psi^*} = \frac{\ee^{\ii\Phi}}{2}\left(\pd{}{\Psi} + \frac{\ii}{\Psi}\pd{}{\Phi}\right)\,.
\end{equation}
From this we see that
\begin{equation}
\pd{f(\psi, \psi^*)}{\psi} = 0 = \pd{f(\psi, \psi^*)}{\psi^*} \ \Leftrightarrow\ \pd{f(\psi, \psi^*)}{\Psi} = 0 = \pd{f(\psi, \psi^*)}{\Phi}\,.
\end{equation}
Thus it will suffice to look at derivatives with respect to the phases of the $\psi_i$. This drastically simplifies the problem, because $f$ and $g$ only depend on $|\psi_i| \equiv \Psi_i$.

Now we need to know how to handle the $\Im$-symbol in the formula. By definition $\Im f = \frac{1}{2\ii} (f - f^*)$, so
\begin{align}
\pd{\Im f(\psi, \psi^*)}{\Phi} =&\, \pd{\Im f}{f}\pd{f(\psi, \psi^*)}{\Phi} + \pd{\Im f}{f^*}\pd{f^*(\psi, \psi^*)}{\Phi} \notag\\
 =&\, \frac{1}{2\ii}\pd{f(\psi, \psi^*)}{\Phi} - \frac{1}{2\ii}\pd{f^*(\psi, \psi^*)}{\Phi} = \Im \pd{f(\psi, \psi^*)}{\Phi}\,.
\end{align}
Here we used that
\begin{equation}
\pd{f^*(\psi, \psi^*)}{\Phi} = \left(\frac{1}{2\ii}\pd{f(\psi, \psi^*)}{\Phi^*}\right)^* = \left(\frac{1}{2\ii}\pd{f(\psi, \psi^*)}{\Phi}\right)^*
\end{equation}
because $\Phi \in \mathbb{R}$. Finally we can state that $\pd{}{\Phi}$ and $\Im$ commute.

For briefness in the calculation also note that
\begin{equation}
\pdn{\psi_i^n}{\Phi_j}{m} = (\ii n)^m \psi_i\delta_{ij}\,,\qquad \pdn{(\psi_i^*)^n}{\Phi_j}{m} = (-\ii n)^m \psi_i\delta_{ij}
\end{equation}
and $\Im (\ii f) = \Re f$.

\subsection{The derivatives of the Poisson-brackets} \label{subsubsec:compPois}

Now we will look at different derivatives of $\lbrace H, N_k\rbrace$ and impose they all vanish. This will lead to a contradiction with our assumptions.

In the computations we will implicitly use the relations of \cref{subsubsec:tech}, but as it would unnecessarily elongate this section, we will not explicitly point out every single step.
\begin{align}
0 = \frac{\partial^3 \lbrace H, N_k \rbrace}{\partial \Phi_1 \partial \Phi_3 \partial \Phi_2} = -8\ii\Re\Big{(}
& Ju_{1k}u_{3k}^*\psi_1^*\psi_1^*\psi_3\psi_2 f^2(\Psi_1^2)f(\Psi_3^2)f(\Psi_2^2) \notag\\
&\,+ J^*u_{1k}^*u_{3k}\psi_1\psi_3^*\psi_3^*\psi_2 f(\Psi_1^2)f^2(\Psi_3^2)f(\Psi_2^2) \Big{)}
\end{align}
\begin{equation} \label{eq:Re1}
\Leftrightarrow\, \Re Ju_{1k}u_{3k}^*\psi_1^*\psi_1^*\psi_3\psi_2 f^2(\Psi_1^2)f(\Psi_3^2)f(\Psi_2^2) = -\Re J^*u_{1k}^*u_{3k}\psi_1\psi_3^*\psi_3^*\psi_2 f(\Psi_1^2)f^2(\Psi_3^2)f(\Psi_2^2)\,.
\end{equation}
\begin{align}
0 = \frac{\partial^4 \lbrace H, N_k \rbrace}{\partial \Phi_1 \partial \Phi_3 \left(\partial \Phi_2\right)^2} = 8\ii\Im\Big{(}
&  Ju_{1k}u_{3k}^*\psi_1^*\psi_1^*\psi_3\psi_2 f^2(\Psi_1^2)f(\Psi_3^2)f(\Psi_2^2) \notag\\
&\,+ J^*u_{1k}^*u_{3k}\psi_1\psi_3^*\psi_3^*\psi_2 f(\Psi_1^2)f^2(\Psi_3^2)f(\Psi_2^2) \Big{)}
\end{align}
\begin{equation} \label{eq:Im1}
\Leftrightarrow\, \Im Ju_{1k}u_{3k}^*\psi_1^*\psi_1^*\psi_3\psi_2 f^2(\Psi_1^2)f(\Psi_3^2)f(\Psi_2^2) = -\Im J^*u_{1k}^*u_{3k}\psi_1\psi_3^*\psi_3^*\psi_2 f(\Psi_1^2)f^2(\Psi_3^2)f(\Psi_2^2)\,.
\end{equation}
\begin{align}
0 = \frac{\partial^4 \lbrace H, N_k \rbrace}{\left(\partial \Phi_1\right)^2 \partial \Phi_3 \partial \Phi_2} = 8\ii\Im\Big{(}
&\,- 2Ju_{1k}u_{3k}^*\psi_1^*\psi_1^*\psi_3\psi_2 f^2(\Psi_1^2)f(\Psi_3^2)f(\Psi_2^2) \notag\\
&\,+ J^*u_{1k}^*u_{3k}\psi_1\psi_3^*\psi_3^*\psi_2 f(\Psi_1^2)f^2(\Psi_3^2)f(\Psi_2^2) \Big{)}
\end{align}
\begin{equation} \label{eq:Im2}
\Leftrightarrow\, 2\Im Ju_{1k}u_{3k}^*\psi_1^*\psi_1^*\psi_3\psi_2 f^2(\Psi_1^2)f(\Psi_3^2)f(\Psi_2^2) = \Im J^*u_{1k}^*u_{3k}\psi_1\psi_3^*\psi_3^*\psi_2 f(\Psi_1^2)f^2(\Psi_3^2)f(\Psi_2^2)\,.
\end{equation}
\begin{align}
0 = \frac{\partial^5 \lbrace H, N_k \rbrace}{\left(\partial \Phi_1\right)^2 \partial \Phi_3 \left(\partial \Phi_2\right)^2} = 8\ii\Re\Big{(}
&\,- 2Ju_{1k}u_{3k}^*\psi_1^*\psi_1^*\psi_3\psi_2 f^2(\Psi_1^2)f(\Psi_3^2)f(\Psi_2^2) \notag\\
&\,+ J^*u_{1k}^*u_{3k}\psi_1\psi_3^*\psi_3^*\psi_2 f(\Psi_1^2)f^2(\Psi_3^2)f(\Psi_2^2) \Big{)}
\end{align}
\begin{equation} \label{eq:Re2}
\Leftrightarrow\, 2\Re Ju_{1k}u_{3k}^*\psi_1^*\psi_1^*\psi_3\psi_2 f^2(\Psi_1^2)f(\Psi_3^2)f(\Psi_2^2) = \Re J^*u_{1k}^*u_{3k}\psi_1\psi_3^*\psi_3^*\psi_2 f(\Psi_1^2)f^2(\Psi_3^2)f(\Psi_2^2)\,.
\end{equation}
From \cref{eq:Re1,eq:Im1,eq:Im2,eq:Re2} we can conclude that
\begin{equation}
\Re Ju_{1k}u_{3k}^*\psi_1^*\psi_1^*\psi_3\psi_2 f^2(\Psi_1^2)f(\Psi_3^2)f(\Psi_2^2) = \Im Ju_{1k}u_{3k}^*\psi_1^*\psi_1^*\psi_3\psi_2 f^2(\Psi_1^2)f(\Psi_3^2)f(\Psi_2^2) = 0
\end{equation}
and thus
\begin{equation}
Ju_{1k}u_{3k}^*\psi_1^*\psi_1^*\psi_3\psi_2 f^2(\Psi_1^2)f(\Psi_3^2)f(\Psi_2^2) = 0\,.
\end{equation}
This however, is in direct contradiction to our assumptions, because basically we assumed every part of this expression to be non-zero in \cref{subsubsec:tech}
\end{appendices}

\newpage
\bibliographystyle{unsrt}
\bibliography{\curPath /references}

\end{document}